%
%     plain-TeX version for world scientific conference proceedings
%
%--------------------------- CUT HERE --------------------------------------
%
%Updated with effect from: 5 Sept 1991
\input epsf.tex
\overfullrule=0pt
\def\np#1#2#3{{\elevenit Nucl. Phys. \/}{\elevenbf B#1} (#2) #3 }
\def\pl#1#2#3{{\elevenit Phys. Lett. \/}{\elevenbf {#1}B} (#2) #3}
\def\prl#1#2#3{{\elevenit Phys. Rev. Lett. \/}{\elevenbf #1} (#2) #3}
\def\physrev#1#2#3{{\elevenit Phys. Rev. \/}{\elevenbf D#1} (#2) #3}
\def\CO{{\cal O}}
\headline={\ifnum\pageno=1\firstheadline\else
\ifodd\pageno\rightheadline \else\leftheadline\fi\fi}
\def\foot#1#2{\footnote{#1}{{\ninerm\baselineskip=11pt
#2}\hfil}}
\def\firstheadline{DOE/ER/40561-189-INT95-00-86 \hfil}
\def\rightheadline{\hfil}
\def\leftheadline{\hfil}
        \footline={\ifnum\pageno=1\firstfootline\else\otherfootline\fi}
\def\firstfootline{\rm\hss\folio\hss}
\def\otherfootline{\hfil}
\font\tenbf=cmbx10
\font\tenrm=cmr10
\font\tenit=cmti10
\font\elevenbf=cmbx10 scaled\magstep 1
\font\elevenrm=cmr10 scaled\magstep 1
\font\elevenit=cmti10 scaled\magstep 1

\font\ninerm=cmr9

%\TagsOnRight
\nopagenumbers
\line{\hfil }
\vglue 1cm
\hsize=6.0truein
\vsize=8.5truein
\parindent=3pc
\baselineskip=10pt
\centerline{\tenbf ELECTROWEAK BARYOGENESIS}
\vglue 1.0cm
\centerline{\tenrm DAVID B. KAPLAN }
\baselineskip=13pt
\centerline{\tenit Institute for Nuclear Theory NK-12, University of
Washington}
\baselineskip=12pt
\centerline{\tenit Seattle, WA 98195, USA}
\vglue 0.8cm
\centerline{\tenrm ABSTRACT}
\vglue 0.3cm
{\rightskip=3pc
 \leftskip=3pc
 \tenrm\baselineskip=12pt%\parindent=1pc
 \noindent
Baryogenesis during the electroweak phase transition is a plausible scenario
for for the origin of matter in the Universe.  Furthermore, it has the
advantage over other scenarios in that one can imagine the much of the physics
involved may be experimentally probed before long.  In the past year a
consensus has developed about the major mechanisms involved.   In this talk I
give an overview of the standard picture,  and discuss briefly the advances
over the past year that suggest electroweak baryogenesis is a robust
phenomenon.
 \vglue 0.8cm }
\line{\elevenbf 1. General Mechanism for Electroweak Baryogenesis \hfil}
\bigskip
\line{\elevenit 1.1.Why Electroweak Baryogenesis is Interesting \hfil}
\smallskip
\baselineskip=14pt
\elevenrm

The quantity we wish to explain is $n_B/s$ --- the ratio of the baryon density
of the Universe to the entropy --- observed to be equal to$^1$
$$n_B/s\simeq (0.8\pm0.2)\times 10^{-10}\eqno(1)$$
Approximately thirty years ago Sakharov outlined three basic requirements for
an explanation of this ratio from microphysics.   These are (1) B violation;
(2) C and CP violation; (3) Departure from thermal equilibrium.   It is
intriguing that all three ingredients are present in the Standard Model (SM)
plus Big Bang: B violation can occur in the SM, as shown by 't Hooft, and both
C and CP violation have been observed.  Departure from thermal equilibrium can
occur due to the expansion of the Universe  in the Big Bang theory: it is
generic  at very early times when the cooling rate of the Universe is more
rapid than particle interactions, but can also occur at late times if there is
a first order phase transition with supercooling$^2$.

The first models for baryogenesis satisfied the Sakharov criteria in Grand
Unified (GUT) theories, exploiting the fact that the GUT scale was not far from
the Planck scale.  Thus baryon violation could exist without being in conflict
with the proton lifetime, and departure from equilibrium could easily result
due to the rapid expansion of the Universe during the GUT epoch.  All such
theories had to involve new sources of CP violation, as Kobayashi-Maskawa CP
violation alone proved to be too small to explain the observed baryon
asymmetry.

Two subsequently discovered effects threaten the viability of GUT-scale
baryogenesis.  The first is inflation, which serves to wash out GUT scale
monopoles, but also washes out any baryon asymmetry produced prior to
inflation.     For GUT scale baryogenesis to occur after inflation requires a
high reheat temperature, which requires a strongly coupled inflaton, which in
turn tends to give large density perturbations inconsistent with structure
formation.  A second problem is that it is now thought that the anomalous
baryon violation of the SM discovered by 't Hooft occurs relatively rapidly at
high temperature$^{2,3}$.  These interactions cause any baryon asymmetry to
equilibrate to a number proportional to B-L, a quantum number preserved by the
SM.   ``New and improved''  GUT scale baryogenesis models must now posess an
effective B-L symmetry that is violated at high energies, and the baryogenesis
must typically involve a scalar field that can ``store up'' B-L during
inflation,  only releasing it after inflation is over.  Such a field can be the
inflaton itself, or something like a squark or slepton field in supersymmetry
(SUSY) models$^4$.   Unfortunately, the only new observable experimental
consequences from such elaborate constructs  are B or  L violation, and even
then the rates are very model dependent and can be adjusted to be out of reach.
 The additional CP violation cannot be observed directly.  In fact, due to
inflation, one can account for a baryon asymmetry within our horizon while
having no net baryon number for the Universe as a whole, thus eliminating the
need for CP violation.

It is remarkable that the SM itself gives rise to baryogenesis, even though it
predicts at best $n_B/s\simeq 10^{-20}$, and perhaps a lot less, depending on
the Higgs mass.  Thus while our very existence can be claimed to be the most
compelling evidence for new physics beyond the SM (perhaps only second to the
existence of gravity!) it is possible that only relatively minor extensions of
the SM at the electroweak scale are required.  One bottleneck for electroweak
baryogenesis (EWB) proves to be inadequate CP violation from the
Kobayashi-Maskawa  mechanism (as was found in the original GUT baryogenesis
models).  Another is the need for departure from thermal equilibrium, which can
only occur at the electroweak epoch only if there is a sufficiently strong
first order phase transition.  As discussed below, this requires a light SM
Higgs, or an extension of the Higgs sector.

If baryogenesis proceeds at the electroweak scale,  we might hope to see new CP
violating effects in terrestrial experiments, such as measurements of the
electron and neutron electric dipole moments, or direct detection of the new
particles and interactions responsible for additional CP violation.
Furthermore, EWB requires there to be a first order phase transition, which
places constraints on the Higgs sector which might testable some day.  A final
argument for considering electroweak baryogenesis is one of economy:  any
theory of baryogenesis needs to posit physics beyond the SM, but for EWB the
additions required are quite minimal.

\bigskip
\line{\elevenit 1.2. The One Basic Mechanism with Four Constraints \hfil}
\smallskip
The standard picture of electroweak baryogenesis assumes that $SU(2)\times
U(1)$ breaks via a first order phase transition, leading to bubble nucleation
and a separation of phases$^2$.  In the symmetric phase, anomalous baryon
violation is occuring rapidly, at a rate/unit volume estimated as
$\Gamma_{sym.}\sim (\alpha_w T)^4$; while in the broken phase the process
occurs due to sphalerons and the rate is estimated to be $\Gamma_{brok.}\sim
M_W^7/(\alpha_w T)^3 exp[-E_{sp}/T]$, where $E_{sp} \simeq 2M_w/\alpha_w$ is
the sphaleron energy$^3$.    In order to produce a baryon asymmetry, it is
necessary that as the bubble of broken phase nucleates and expands, particle
interactions with the bubble wall somehow produce a baryon excess in the
symmetric phase, where B violation is rapid.
However, this baryon asymmetry will be subsequently destroyed, unless the B
violation rate in the broken phase be extremely small.  That is, the phase
transition must be sufficiently strong so that sphalerons are heavy and play no
role in baryogenesis.

But how does the baryon asymmetry actually come about?    Ref.~6 was the first
model proposed for EWB, and it outlined the four basic requirements that have
been generally incorporated into other models:
\smallskip
\item{i.} Particle interactions with the bubble wall lead to a pile-up of
particles in front of the advancing bubble wall.  This is how the departure
from thermal equilibrium translates into a distortion of the particle
distribution functions, and it requires  some particles with strong coupling to
the Higgs field in the bubble wall.
\smallskip
\item{ii.}
These particles being swept in front of the bubble carry net $SU(2)$
left-handed doublet number, which biases anomalous baryon violation in the
direction of reducing this excess.
This requires sufficient CP violation, since doublet number is CP-odd.
\smallskip
\item{iii.}  Each anomalous event that reduces the doublet number also produces
three units of baryon number.   Once baryon number is produced in the symmetric
phase, it is overtaken by the bubble wall and enters the broken phase, where B
violation is effectively absent.  This requires a sufficiently strong first
order phase transition, as mentioned above, which in turn places constraints on
the Higgs sector.

\smallskip
\item{iv.} The baryon violation rate is proportional to $\alpha_w^4 \sim
10^{-6}$; given enough time to equilibrate, this rate would drop out of the
final value for the baryon asymmetry, but otherwise the resultant baryon
asymmetry will be reduced by this factor.  Furthermore, $n_B/s\propto 1/g_*
\simeq 10^{-2}$, where $g_*$ counts degrees of freedom at the critical
temperature.  Thus there is only room for an additional factor of $10^{-2}$
from the CP violating angle and any inefficiency of the dynamical mechanism.
This proves to be too severe a constraint to satisfy unless there is an enhance
of the form of a ratio of time scales:  the excess $SU(2)$ doublets must spend
sufficiently long time in the symmetric phase before being over taken by the
bubble wall.  This  requires significant transport of doublet number into the
symmetric phase.  Even then, one finds that  CP violation must occur at the
$10^{-3}$ level or larger in most models.
\smallskip

\noindent
The model of Ref.~6  consisted of the SM with the addition of three
right-handed neutrinos and a singlet scalar (the ``singlet majoron
model$^7$''). The right-handed neutrinos had large Yukawa couplings, satisfying
constraint (i); the neutrino mass matrix contained large CP violating angles,
satisfying constraint (ii);  with the addition of the singlet majoron field,
the weak phase transition can be quite strongly first order for a range of
parameters, satisfying (iii);  finally, significant charge transport of
left-handed doublet number into the symmetric phase occurred in the form of low
energy left-handed neutrinos (which interact weakly), satisfying (iv).  The
model is testable in that it predicts a mass for the $\nu_\tau$ in the range
5-30 MeV, as well as an extended Higgs sector with an extra complex scalar.
The constraint on the $\nu_\tau$ mass comes from a generic feature of EWB
models that the new  CP violation must be sizable.

I now turn to discussing the current status of meeting our four requirements in
general extensions of the standard model.  As much of the details are contained
the the review Ref.~8, I will simply summarize results, and give particular
emphasis to advances of the past year.

\vglue 0.6cm
\line{\elevenbf 2. Current Status of the Four Constraints \hfil}
\vglue 0.4cm

\bigskip
\line{\elevenit 2.1.  Significant Higgs--Particle interaction. \hfil}
\smallskip
Heavy particles are required, since only they will interact strongly with the
Higgs field in the bubble wall.   This constraint is no challenge to meet:
either the top quark is the engine that drives EWB; or exotic heavy particles
such as right-handed neutrinos, squarks, or higgsinos$^9$.  In a SUSY model
with large $\tan\beta$, even the $b$ quarks and $\tau$ leptons might
participate$^{10}$.

\bigskip
\line{\elevenit 2.2.  Sufficient CP Violation \hfil}
\smallskip
{\elevenbf Old Common Wisdom}:
Since we see CP violation in the kaon system, it is of great interest to know
whether  the same CP violation is what gave rise to matter in the Universe. A
rough estimate suggests that since CP violation in the SM would vanish if any
two quarks of the same charge had the same mass, the dimensionless measure of
CP violation should be proportional to  $G_F^6 \prod_{i> j} (m_{u_i}^2
-m_{u_j}^2)(m_{d_i}^2-m_{d_j}^2)\simeq 10^{-20}$. An explicit computation of
GUT scale baryogenesis in  minimal $SU(5)$ confirms such a suppression$^{11}$.
If this counting is correct, then clearly additional sources of CP violation
are requred to explain (1).  Possible additional sources of CP violation
include extra Higgs doublets; massive neutrinos; and SUSY.  As mentioned above,
typical CP violating angles have to be $10^{-3}$ or greater, which usually has
experimental implications.  In the two-Higgs and SUSY models, the EDM of the
electron is within two orders of magnitude of present bounds; in the singlet
Majoron model, $\nu_\tau$ must be within an order of magnitude of the present
bound.

{\elevenbf New CW}:
Recently$^{12}$ Shaposhnikov and Farrar challenged the above reasoning, which
is admittedly naive.  There are other, larger, plausible measures of CP
violation.  And after all, $\epsilon\gg10^{-20}$ in the kaon system.   Ref.~12
claims that in fact KM CP violation can be large enough to explain the baryon
asymmetry; however   in the past year there have been two challenges to their
work, however, which conclude that the naive estimate is correct and that KM CP
violation would lead to a baryon asymmetry too small by perhaps ten orders of
magnitude$^{13}$.  Personally, I find the later work convincing, and believe
that new CP violation must be added to any theory of baryogenesis.

\bigskip
\goodbreak
\line{\elevenit 2.3.  A Strong First-Order Phase Transition. \hfil}
\smallskip
{\elevenbf Old CW}: There has been extensive work on the nature of the
electroweak phase transition, and there is general agreement that for the SM
with one Higgs doublet, the transition is strongly first order in the limit
$M_H \ll M_W$ ({\elevenit e.g,} as in the Coleman-Weinberg scenario) while
being second order in the $M_H\gg M_W$ limit.  Perturbative calculations break
down for $M_H\simeq M_W$.
The maximum value for the Higgs mass where baryon asymmetry produced during the
phase transition is not subsequently destroyed in the broken phase was thought
to occur at about $M_H\le 45$ MeV, which is experimentally  excluded. It was
also thought that extended Higgs sectors, such as 2-Higgs models and the
singlet Majoron model are still viable candidates for a sufficiently strong
first order phase transition. A notable exception is the minimal SUSY model,
which is only viable should there be a relatively light squark.

{\elevenbf New CW}:  There have been two recent advances in studies of the
electroweak phase transition. One is an novel application of the $\epsilon$
expansion$^{14}$, which suggests that (i) The phase transition is more strongly
first order than perturbation theory suggests, but that (ii) the baryon
violation rate in the broken phase is faster than previously calculated. That
is, for a given Higgs mass, $\langle H\rangle$ is larger but $g$ is smaller in
the broken phase, in comparison with perturbative results.  The
$\epsilon\rightarrow 1$ limit is not sufficiently under control to give a
precise Higgs bound, however.

A second advance has been in lattice simulations of the electroweak phase
transition$^{15}$.    These simulations also tend to suggest that
nonperturbative effects are important for physically interesting Higgs masses,
and that the phase transition tends to be more strongly first order than found
in perturbation theory.  For example, one simulation found  that for a Higgs
mass of 80 GeV,  $\langle H\rangle/T_c = 0.68$, as opposed to $0.3$ from
perturbation theory.   Thus the upper bound of 45 MeV on the Higgs mass for EWB
is probably too conservative, although I have not seen any definitive result
replace it.  Perhaps the one doublet model is still barely viable so far as the
phase transition constraint goes.

\bigskip
\line{\elevenit 2.4 Transport \hfil}
\smallskip

{\elevenbf Old CW}:
The issue here is one of time scales, and has been a major point of discussion
in the literature over the past year.  The anomalous baryon violation rate/unit
volume in the symmetric phase is parametrized as $\Gamma= \kappa \alpha_w^4
T^4$.  The parameter $\kappa$ is a fudge factor to parametrize our ignorance;
there has been some lattice evidence for it to be $\CO(1)$ $^{16}$.    For a
crude estimate of how much baryon number we can make during the electroweak
phase transition, assume that there is a region in front of the wall with an
excess  $SU(2)$ fermion doublets density $\delta n_d$, and that it extends a
length $\ell$ in front of the bubble wall.  As the bubble wall sweeps through
all of space with a velocity $v_w$,  each point in space is in the doublet rich
region for a time $\delta t \sim \ell/v_w$. During that time the baryon
production rate is given roughly by $\Gamma \delta n_d/T^3$.  The resultant
value for $n_b/s$ is then given by
$$n_B/s\sim \left({\kappa \alpha_w^4 \theta_{CP}\over g_*}\right) (\ell T_c)
\left({\delta n_d \over \theta_{CP} v_wT_c^3}\right)\ , \eqno(2)$$
where the excess doublet density  $\delta n_d$ is proportional to CP violating
parameter $\theta_{CP}$, and is  typically  proportional to the wall velocity
$v_w$ as well.  The above formula assumes $\ell$ is short enough that baryon
violation does not have enough time to fully equilibrate --- otherwise the B
violation rate would drop out of the formula\foot{$^*$}{Eq.~2 is for
illustrative purposes only --- much more sophisticated analyses have been
discussed in the literature.}.

Since the first term in Eq. (2) is $\CO(10^{-10})$ for an angle
$\theta_{CP}\sim 10^{-2}$, evidently a sizable value for the parameter $\ell$
is crucial for making enough baryon number.  If the bubble wall is thin, then
particles bounce off it and can be reflected for a long way into the symmetric
phase.  The method for computing the flux of weak doublets reflecting from the
bubble walls was presented in Refs.~6,17,18  for the both   the singlet majoron
model and the two Higgs model with a thin bubble wall and top quark reflection.
 Transport properties were considered in Ref.~17, where  a Monte Carlo
calculation showed a significant ``snowplow'' effect: lefthanded top quarks
were pushed along in a region ranging from 20 to 100 thermal units in front of
the bubble wall (Fig. 1).  Such a model was shown to easily accomodate the
desired baryon asymmetry (1) with $\theta_{CP} = 10^{-2} - 10^{-3}$.
\bigskip
\topinsert{\centerline{\epsfxsize=4.0in\epsfbox{
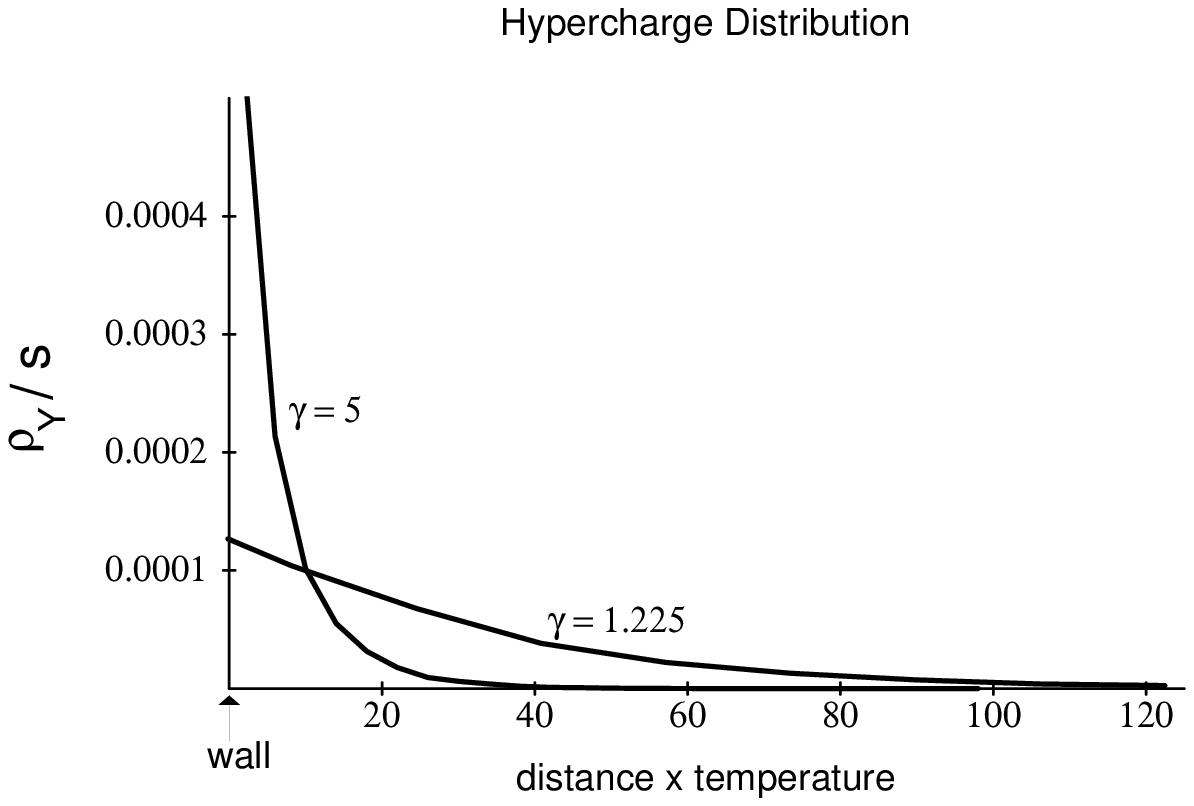}}
\smallskip
\centerline{\vbox{\rightskip=3pc
 \leftskip=3pc
 \tenrm\baselineskip=12pt%\parindent=1pc
 \noindent
 Fig. 1. Axial top density in the two Higgs model with wall width $1/T_c$
(``thin wall''), top quark mass $2T_c$, and wall velocities $v_w = 0.98$
9$\gamma=5$) and $v_w=1/\sqrt{3}$ ($\gamma=1.225$).  Note that the baryon
biasing doublet density extends far into the symmetric phase. From
Ref.~17.}}}\endinsert
\bigskip
For broad bubble walls --- thicker than the mean free path of particles --- the
process was supposed to progress quasi-statically within the bubble wall, where
the time dependent  Higgs field acted like a chemical potential for left handed
doublet number.  This scenario --- dubbed ``spontaneous baryogenesis'' ---
appeared to work, but only barely, since the factor $\ell$ in Eq. (2) was
effectively the wall width, and there were various suppression factors$^{19}$.

{\elevenbf New CW}.  While the thin wall scenario remains unchanged, there has
been a lot of work done recently for phase transitions with a thick bubble
wall.  Dine and Thomas pointed out a number of suppression factors in the thick
wall scenario$^{20}$.  The basic problem they found was that the CP violating
effects gave rise to a doublet density in the region of the wall where the
Higgs field was large, while baryon violation only occured where the Higgs
field was small.  In the adiabatic approximation, there was no transport  and
little overlap between the two regions; the baryon asymmetry was estimated to
be at best two orders of magnitude too small.  A second objection to the
adiabatic treatment for thick walls was made by Giudice and Shaposhnikov, who
pointed out that QCD anomalous events wanted to turn left handed doublets into
right handed ones, which in thermal equilibrium completely eliminated the weak
doublet density that was biasing baryon production$^{21}$.  Finally there was a
preprint by Joyce, Prokopec and Turok which pointed out that diffusion had not
been properly accounted for in the adiabatic case, and that it would further
suppress baryon violation$^{22}$.

%\bigskip
\topinsert{\centerline{\epsfxsize=4.0in\epsfbox{
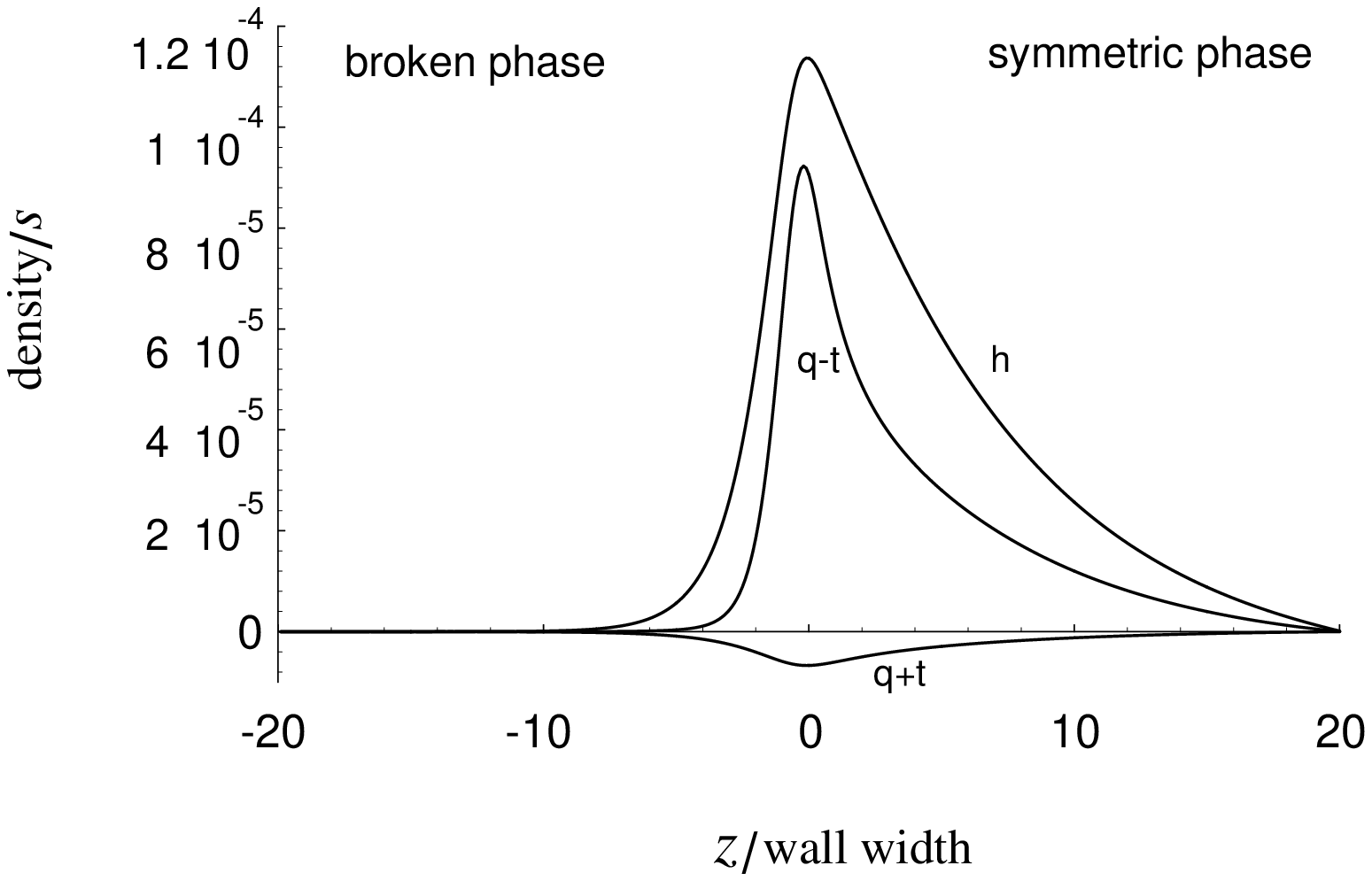}}
\smallskip
\centerline{\vbox{\rightskip=3pc
 \leftskip=3pc
 \tenrm\baselineskip=12pt%\parindent=1pc
 \noindent
 Fig. 2. Particle asymmetry densities as a function of wall width, computed
from a driven diffusion equation.  $z=0$ corresponds to the center of the wall.
The wall width was taken to be $10/T_c$ and the wall velocity $v_w = 0.1c$.
The particle densities $h$, $q$ and $t$ refer to Higgs particles, $t_L+b_L$,
and $t_R$ respectively, and were shown to specify the 3 family system
completely.  From Ref.~23.  }}}\endinsert
\bigskip

In fact it was shown in Ref.~23 that diffusion {\elevenit was} important and
that rather than being detrimental, it helped alleviate the other
problems\foot{$^*$}{Some similar conclusions are reached in Refs.~24.}.
Diffusion allows doublet density produced well within the bubble wall to
venture out into the symmetric phase where baryon violation is rapid,
eliminating the Dine-Thomas objection.    In Ref.~23 diffusion equations were
solved in the presence of the bubble wall for the two Higgs model with maximal
CP violation;  the results for one set of parameters is shown in Fig.~2.  Note
that doublet densities diffuse over 100 thermal lengths, so that Fig.~2 looks
not unlike Fig.~1 which was computed in the thin wall approximation by Monte
Carlo.

Fig.~3 shows how the Dine-Thomas objection is evaded.  We plot the baryon
asymmetry that results as a function of $z_{co}$, the point in the wall where
one assumes that baryon violation is cut off.  Dine and Thomas argued that
$z_{co}\simeq 2.5$\foot{$^*$}{They actually express the cutoff in terms of a
value for $\langle H\rangle$, which translates to $z_{co}\simeq 2.5$ in the
model considered.}.  Without diffusion, we find their suppression (dotted
line).  However, including diffusion makes the final result very insensitive to
details of baryon violation within the bubble wall, and we see that a CP
violating angle of $10^{-2}-10^{-3}$ once again provides a sufficient baryon
asymmetry for the Universe.

%\bigskip
\topinsert{\centerline{\epsfxsize=4.0in\epsfbox{
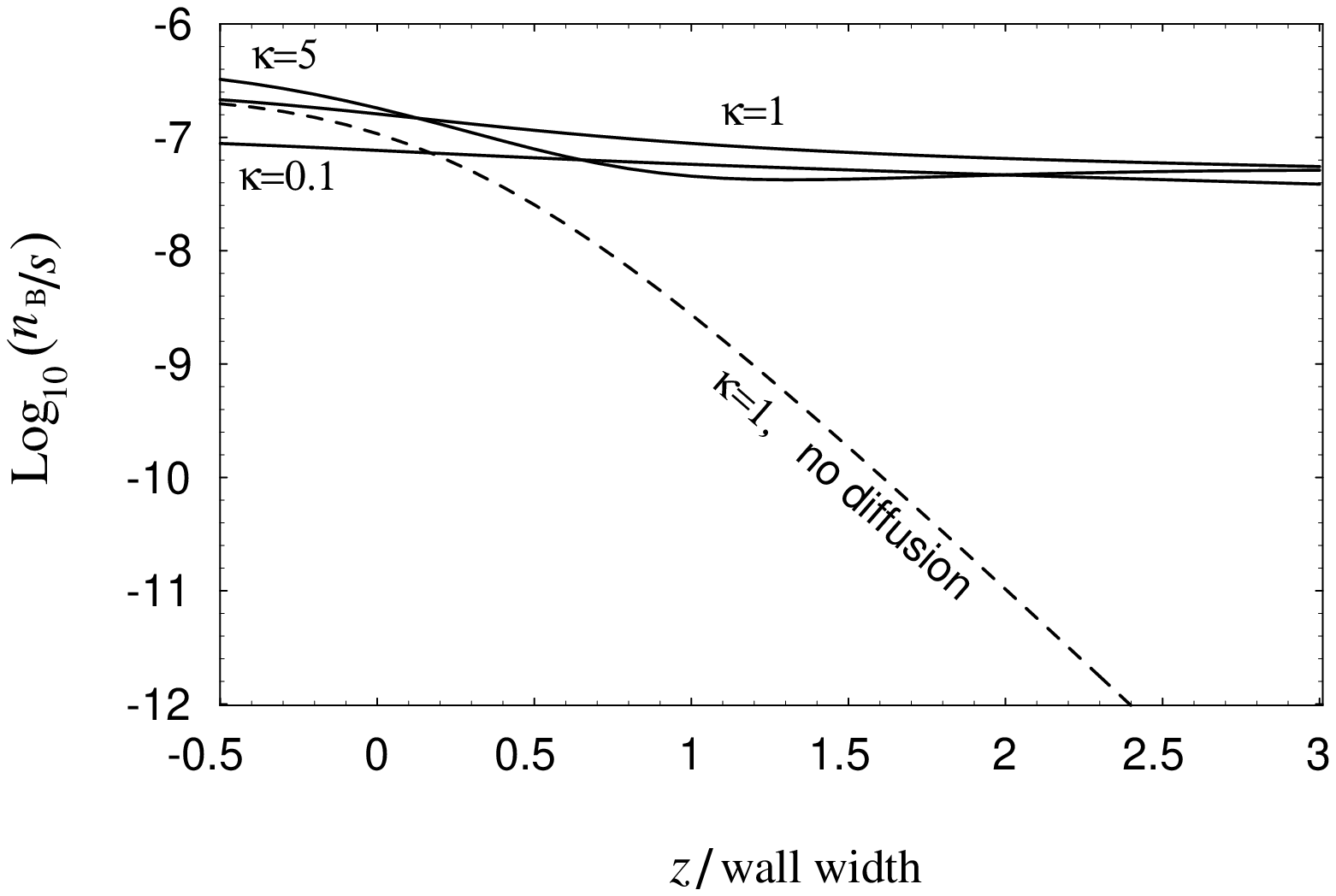}}
\smallskip
\centerline{\vbox{\rightskip=3pc
 \leftskip=3pc
 \tenrm\baselineskip=12pt%\parindent=1pc
 \noindent
 Fig.~3. Resultant baryon asymmetry as a function of where in the wall baryon
violation is assumed to be cut off, for various values of $\kappa$. Note the
importance of diffusion for $z_{co} > 0.5$, and the insensitivity to $\kappa$.
{}From Ref.~23.   }}}\endinsert
\bigskip
\goodbreak
Fig.~3 reveals an additional bonus.  The  diffusion equations properly account
for the finite rates of interactions, and one finds that particles are not in
front of the wall long enough for strong sphalerons to completely eliminate the
left handed doublet density.  Rather, there is a competition between the
$SU(2)$ and $SU(3)$ anomalies and the final baryon asymmetry is roughly
proportional to the ratio of their rates, which makes the final answer quite
insensitive to (unknown) nonperturbative physics.  Thus the Giudice
Shaposhnikov observation leaves EWB not only viable, but less model dependent.

The analysis of Ref.~23 made a number of approximations that are invalid for
various extreme limits of the wall velocity or the wall width.  A more detailed
analysis is found in Refs.~25, which for a wide range of parameters are in
general agreement with the analysis described above.  Overall, the new CW for
this section must be described as quite optimistic for EWB.

\vglue 0.6cm
\line{\elevenbf 3. Conclusions... and What Next?  \hfil}
\vglue 0.4cm

Until last year, there were two apparently unrelated regimes for EWB -- the
thin wall (``nonadiabatic'', ``charge transport'', or ``nonlocal'') regime
which seemed to work well, but only applied to certain types of Higgs sectors
--- and the thick wall (``adiabatic'', ``spontaneous baryogenesis'',
``local'') regime, which had a pretty theory behind it, but appeared to be on
the edge of viability.  The picture that has emerged recently is much more
unified, as seen in the comparison of Figs. 1,2:  all EWB is nonlocal, and
charge transport plays a crucial role in giving the relatively slow $SU(2)$
anomaly time to make enough baryon number.  It seems clear at this point that
EWB has come of age and is a viable theory; the question remains, is it right?

In principle, knowing all of the extra particles and interactions in an
extension of the SM will allow us to compute the baryon asymmetry generated at
the electroweak phase transition, perhaps to an order of magnitude.  Such a
wealth of knowledge seems far off, however.
So far  experimental signals have appeared to be quite model dependent,
although having enough CP violation has constrained all models seen to date to
have experimental consequences within two orders of magnitude of current
measurements -- either the EDM of the electron, or the $\nu_\tau$ mass.  It
would be interesting to see if there are any model-independent phenomenological
predictions.  It would also be interesting to see SUSY models analyzed in depth
for their ability to create baryon number at the weak phase transition.

Needless to say, this talk is far from a complete review of EWB, but rather an
admittedly biased view of what I consider the most interesting recent
developments.

 \vglue 0.6cm
\line{\elevenbf 4. Acknowledgements \hfil}
\vglue 0.4cm
I wish to acknowledge my collaborators on EWB, Andrew Cohen and Ann Nelson.
This work was supported in part by DOE grant DOE-ER-40561, NSF Presidential
Young Investigator award PHY-9057135, and by a grant from the Sloan Foundation.
\vglue 0.6cm
\line{\elevenbf 6. References \hfil}
\vglue 0.4cm

\medskip
\item{1.} E.W. Kolb,M.S. Turner, {\elevenit The Early Universe}
(Addison-Wesley, New York, 1990)
\item{2.} V.A.  Kuzmin, V.A.   Rubakov, M.E. Shaposhnikov, Phys. Lett.
{\elevenbf 155B} (1985) 36
 \item{3.} P. Arnold,   L. McLerran, Phys. Rev. {\elevenbf D36} (1987) 581;
{\elevenbf D37} (1988)1020
 \item{4.} I. Affleck, M. Dine, Nucl. Phys. {\elevenbf B249} (1985) 361; H.
Murayama, Phys. Lett. B322 (1994) 349; K. Olive, hep-ph/9404352
 \item{5.} A.G. Cohen, D.B. Kaplan, \pl{199}{1987}{251}; \np{308}{1988}{913}
 \item{6.} A.G. Cohen, D.B. Kaplan, A.E. Nelson,  \pl{245}{1990}{561};
\np{349}{1991}{727}
 \item{7.}  Y. Chikashige, R.  Mohapatra, R.  Peccei, \pl{98}{1981}{265}
 \item{8.} A.G. Cohen, D.B. Kaplan, A.E. Nelson, Annu. Rev. Part.  Nucl. Sci.
{\elevenbf 43} (1993) 27
 \item{9.} A.G. Cohen,  A.E. Nelson, \pl{297} {1992}{111}
 \item{10.} M. Joyce, T. Prokopec, N. Turok, \pl{338}{1994}{269}
 \item{11.}  S.M. Barr, G. Segre,  H.A. Weldon, \physrev{20}{1979}{2494}
 \item{12.} G. Farrar, M. Shaposhnikov, \prl{70}{1993}{2833}, \prl
{71}{1993}{210}; hep-ph/9305275; hep-ph/9406387
 \item{13.} M.B. Gavela {\elevenit et al.}, \np{430}{1994}{345};
\np{430}{1994}{382};
P. Huet, E. Sather, \physrev{51}{1995}{379}
 \item{14.} P. Arnold, L. Yaffe,  Phys. Rev. {\elevenbf D49} (1994) 3003
\item{15.} For a summary, see Z. Fodor, hep-lat/9503014
 \item{16.} J. Ambjorn, M. Laursen, M. Shaposhnikov, \pl{197}{1989}{49};  J.
Ambjorn,  T. Askaard, H. Porter,  M.  Shaposhnikov, \pl{244}{1990}{479};
\np{353}{1991}{346}
 \item{17.} A.E. Nelson, D.B. Kaplan,  A.G. Cohen  \np{373}{1992}{453}
 \item{18.} K. Funakubo {\elevenit et al.}, hep-ph/9405422; hep-ph/9407207;
J.M. Cline, \pl{338}{1994}{263}
 \item{19.}  A.G. Cohen, D.B. Kaplan, A.E. Nelson, \pl{263}{1991}{86}
 \item{20.} M. Dine, S. Thomas, \pl{328}{1994}{73}
 \item{21.} G. Giudice, M. Shaposhnikov, \pl{326}{1994}{118}
 \item{22.} M. Joyce, T. Prokopec, N. Turok,  hep-ph/9401351
 \item{23.} A.G. Cohen, D.B. Kaplan, A.E. Nelson,  \pl{336}{1994}{41}
 \item{24.} M. Joyce, hep-ph/9406356; D. Comelli, M. Pietroni, A. Riotto,
hep-ph/9406369
 \item{25.} M. Joyce, T. Prokopec, N. Turok, hep-ph/9410281;  hep-ph/9410282
\vfill \eject
\bye